# A Computational Approach to Bisimulation of Hybrid Dynamical Systems

Babak Tavassoli[1]

*Abstract—* The problem of finding a finite state symbolic model which is bisimilar to a hybrid dynamical system (HDS) and has the minimum number of states is considered. The considered class of HDS allows for discrete-valued inputs that only affect the jumps (events) of the HDS. Representation of the HDS in the form of a transition system is revisited in comparison with prior works. An algorithm is proposed for solving the problem which gives the bisimulation with the minimum number of states if it already exists and also a parameter of the algorithm is properly tuned. There is no need for stability assumptions and no time discretization is applied. The results are applied to an example.

## I. INTRODUCTION

An increasing number of large and complex systems are serving our societies enabled by the technological advances. These new systems do not fit into the classical modeling frameworks anymore. There are various modeling issues in such systems. Some examples are the problem of handling the subsystem interconnections in a networked system, the problem of hiding details to obtain a macro-scale model of a large system, scalable hierarchical design, fault and failure handling in complex systems, security of such systems and interaction of phenomena with different types of mathematical models [1, 2, 3]. The most common form of the later issue is presence of both discrete and continuous states that interact with each other dynamically. Discrete state dynamical systems can be modeled by automaton or Petri net [4]. On the other hand, continuous state dynamical systems are basically modeled by differential equations. A combination of such two systems results in a hybrid state dynamical system or briefly a hybrid dynamical system (HDS) [5, 6]. An HDS can show complex behaviors that are not witnessed in other classes of dynamical systems.

There are several approaches to HDS analysis and control that can be divided into two main categories. The first category is based on the tools and ideas from the control theory. The most important such tools are the Lyapunov and small gain theorems [7, 8]. Other techniques are also applied such as the optimal control based on Hamilton-Jacobi equation [9, 6, 10], dynamic programming for discrete-time stochastic HDS [11] and model predictive control [12]. The second category is based on the ideas and tools from the computer science that try to extend the analysis techniques for discrete systems like the model checking to the HDS. Basically, this is achieved by finding a simplified finite state automaton with the same output space of the HDS such that we can compare the behaviors (output sequences) of the two systems. A finite state system that has the same behavior as the HDS is referred to as a *symbolic model* [13]. Obtaining a suitable symbolic model is an important step toward solving the various practical problems in regard with the HDS. These include safety analysis, reachability analysis and control problems [14, 15, 16]. A symbolic model may be either an *abstraction* or a *bisimulation*. An important distinction is that the behavior of an abstraction only includes the behavior of the HDS, but in the case of bisimulation the two behaviors are exactly the same. Hence, a bisimulation is always more accurate compared to an abstraction, because the abstraction generates behaviors that the original HDS cannot generate. However, it is known that an HDS may fail to have a bisimulation. The existence of finite bisimulations is guaranteed for classes of HDS such as the timed automata, rectangular automata [27, 24]. More recently, existence of a bisimulation is formulated as the O-minimality condition [25], although it is not a necessary and sufficient condition. There is an algorithm for obtaining a bisimulation commonly known as the "bisimulation algorithm" which terminates and gives the bisimulation of an HDS if it exists (Algorithm 8.1 in [13]). However, this algorithm is based on set operations on infinite sets which is not computationally practical. Hence, the bisimulation problem is still an open problem practically and many of the works just try to obtain more accurate abstractions or approximate bisimulations [17, 22]. The notion of approximate bisimulation loosens the notion of bisimulation and can be regarded as an alternative to abstraction. Given some convergence properties of a continuous dynamical system, its reachable set can be approximated by the reachable set of an approximately bisimilar system [17, 18, 13, 19]. A drawback is that the size of the state space of the approximate bisimulation unboundedly increases by decreasing the desired reachability approximation error. Among the symbolic approaches to the HDS problems, it is quite common to consider restricted forms of the control input due to the existing complexities, such as inputs that are finite-valued and constant during intervals [17, 23, 20, 21] or even eliminating the inputs for mere verification [24, 25, 26]. In the existing results on computation of symbolic models, the focus is on the abstraction problem which is easier to solve [19, 28, 29, 22, 32]. Even in the case of abstraction problem, computational costs are high for example due to the required reachablity analysis [30, 26, 22, 11]. In some other results, the reachability analysis is avoided but the costs of computations are still high due to the complex nature of the problem [17, 21]. Abstraction of discrete-time systems is less complex and has gained more attention since a part of the abstraction process which is sampling over time is already performed [23, 31, 11, 22, 19, 17].

In this work we consider the problem of obtaining a bisimulation for an HDS through a computationally feasible method. An algorithm is proposed for computing the bisimulation with minimum number of states. The method is based on partial computation of the HDS behaviors. The proposed algorithm terminates in a finite number of steps which depends on a real-valued parameter $\eta$. If there exists a finite bisimulation, then there exists a value of $\eta$ such that the algorithm gives the finite symbolic model with minimum number of states which is bisimilar to the HDS. However, the right value of $\eta$ cannot be determined before execution of the algorithm. This difficulty is related to undecidability of verification problems for an HDS in the general case [25]. The algorithm needs to be executed with different values of $\eta$ to increase the confidence in correctness of the solution. However, the correct solution when obtained is basically different from the existing results such as the approximate bisimulations [17, 13] whose number of states rapidly increases with the desired accuracy. Finite length behaviors have been also used in [20] to obtain abstractions for a discrete-time HDS without addressing the computation of behaviors. Computational issues for the purpose of abstraction were studied in later works such as [22]. Studying symbolic models based on behaviors appears in other works also such as in [25]. It is remarkable that in contrast to some existing results mentioned above, we do not require any stability or convergence property of the HDS. Also, we do not apply time discretization which is a source of inaccuracy in HDS modeling (e.g. losing the exact time of events). Our definition of HDS is taken from [13] which allows for finite-valued inputs acting on jumps.

The paper is organized as following. Preliminaries are presented in section two. The HDS is represented as a transition system in a form that is suitable for our objective in section three. The required analysis is presented in section four and the algorithms for computation of the bisimulation are presented in section five. A brief example is provided in section six and conclusions are made at the end.

## II. PRELIMINARIES

*Notation*: In the following, $\mathbb{R}$ is the set of real numbers, $\mathbb{R}^+$ is the set of non-negative real numbers and $\mathbb{Z}^+$ is the set of non-negative integers. The Euclidian norm of a vector $\xi \in \mathbb{R}^n$ is denoted by $\|\xi\|$. The boundary of a set $M$ in a metric space is denoted by $\partial M$. For a set $A$, its cardinality is denoted by $|A|$ and the set of all subsets of $A$ is denoted by $2^A$. For a mapping $g: D \to R$ and a set $A \subseteq D$ we define $g(A) = \{r \in R \mid \exists d \in A: r = g(d)\}$. For an equivalence relation $Q$ on a set $Z$, the equivalence class that contains $z \in Z$ is denoted by $[z]_Q$ and the set of all such equivalence classes is denoted by $Z/Q$. For a set $X$ and a sequence of elements $x = \{x_i \in A: 0 \leq i \leq n\}$, the length of $x$ is defined as $|x| = n$. For a sequence $x$, a subsequence of $x$ is another sequence $x'$, if there exists $m \leq |x| - |x'|$ such that $x_{i+m} = x'_i$ for every $0 \leq i \leq |x'|$ and we write $x' = x_{m:n}$ with $n = m+|x'|$ to indicate this relationship. Concatenation of two sequences $x'$, $x''$ denoted by $x = [x'\ x'']$ is defined as $x_i = x'_i$ for $0 \leq i \leq |x'|$ and $x_{i+|x'|+1} = x''_i$ for $0 < i \leq |x''|$. For a closed set $A \subseteq \mathbb{R}^n$ and smooth vector field $f: A \to \mathbb{R}^n$, with $\xi: \mathbb{R} \to \mathbb{R}^n$ satisfying the differential equation $d\xi(t)/dt = f(\xi(t))$ with $\xi(0) = \xi' \in A$, we indicate by $\xi(t) = \mathcal{O}$ the situation in which $\xi$ exits from $A$ at least once during $[0,t)$. More precisely, there exists $t' \in [0,t)$ such that $\xi(t') \in \partial A$, $\lim_{\varepsilon \to 0} \xi(t') + \varepsilon f(\xi(t')) \notin A$ (it is noticed that $\xi$ is undefined outside of $A$). Based on this notation, we define the following functions:

- Transverse time: $\theta_f: A \to \mathbb{R} \cup \{\infty\}$

    $\theta_f(\xi') = \inf \{\{t' \mid \xi(t') = \mathcal{O}\} \cup \infty\}$

- Transition function: $\Phi_f: A \times \mathbb{R} \to \mathbb{R}^n \cup \{\mathcal{O}\}$

    $\Phi_f(\xi', t) = \begin{cases} \xi(t) & 0 \leq t \leq \theta_f(\xi') \\ \mathcal{O} & \text{otherwise} \end{cases}$

- Transverse point: $\Psi_f : \mathbb{R}^n \to \mathbb{R}^n \cup \{*\}$

$$\Psi_f(\xi') = \begin{cases} \Phi_f(\xi', \theta_f(\xi')) & \theta_f(\xi') < \infty \\ * & \theta_f(\xi') = \infty \end{cases}$$

According to the definition of $\Psi_f$, symbol $*$ indicates that $\xi$ never exits from $A$.

## A. Transition Systems

The definitions and theorem in this part are from [13].

*Definition 1*: A *transition system* is a quintuple $S = (X, U, \to, Y, H)$ consisting of:
- a set of states $X$.
- a set of inputs $U$.
- a transition relation $\to \subseteq X \times U \times X$.
- a set of outputs $Y$.
- an output map $H: X \to Y$.

A transition $(x, u, x') \in \to$ describes the untimed evolution of the state from $x \in X$ to $x' \in X$ under the effect of $u \in U$ which is also written as $x \xrightarrow{u} x'$. The output $y \in Y$ is a partial observation of the state $x$. If $X$ is a finite set, then $S$ is said to be a finite transition system. Otherwise, $S$ is an infinite transition system. If $Y = X$, and $H$ is the identity mapping, we briefly write $S = (X, U, \to)$.

A sequence of state transitions $x_0 \xrightarrow{u_0} x_1 \xrightarrow{u_1} \ldots x_{n-1} \xrightarrow{u_{n-1}} x_n$ is an *internal behavior* of length $n$ starting from $x_0$. The corresponding sequence of outputs $y_i = H(x_i)$, $0 \leq i \leq n$ forms an *external behavior* of length $n$. Following the notation in [13], if $n$ is not bounded then the behavior is said to be infinite. An internal behavior of length $n$ is maximal if for every $x' \in X$, $u \in U$ we have $(x_n, u, x') \notin \to$. A transition system is said to be *deterministic* if $x \xrightarrow{u} x'$, $x \xrightarrow{u} x''$ implies that $x' = x''$. We say that $S$ is *non-blocking* if for every $x \in X$ there exists $(x, u, x') \in \to$. We say that $S$ is *output deterministic* if $x \xrightarrow{u'} x'$, $x \xrightarrow{u''} x''$ and $H(x') = H(x'')$ imply that $x' = x''$.

*Definition 2*: Considering two transition systems $S_a = (X_a, U_a, \to_a, Y_a, H_a)$ and $S_b = (X_b, U_b, \to_b, Y_b, H_b)$, we say that $S_a$ and $S_b$ are bisimilar if $Y_a = Y_b$ and there exists a relation $\pi \subseteq X_a \times X_b$ such that for every $(x_a, x_b) \in \pi$ we have
- $H_a(x_a) = H_b(x_b)$.
- $(x_a, u_a, x'_a) \in \to_a$ implies the existence of $(x_b, u_b, x'_b) \in \to_b$ such that $(x'_a, x'_b) \in \pi$.
- $(x_b, u_b, x'_b) \in \to_b$ implies the existence of $(x_a, u_a, x'_a) \in \to_a$ such that $(x'_a, x'_b) \in \pi$.

If $S_a$ and $S_b$ are bisimilar, we can say that $S_b$ is a bisimulation of $S_a$ (and vice versa). The relation $\pi$ is denoted as *bisimulation relation*.

Bisimilarity is a form of equivalence between transition systems because bisimilar transition systems can generate exactly the same set of external behaviors. A transition system can be reduced to a smaller system as below.

*Definition 3*: Let $S = (X, U, \to, Y, H)$ be a system and let $Q$ be an equivalence relation on $X$ such that $(x, x') \in Q$ implies $H(x) = H(x')$. The quotient of $S$ by $Q$, denoted by $S/Q$, is the system $(X/Q, U, \to_Q, Y, H_Q)$ with
- $x_Q \xrightarrow{u}_Q x'_Q$ if there exists $x \xrightarrow{u} x'$ with $x \in x_Q$, $x' \in x'_Q$
- $H_Q(x_Q) = H(x)$ for some $x \in x_Q$.

A quotient system is also denoted as a symbolic model. A symbolic model is of particular interest if it is also a bisimulation according to the following theorem.

*Theorem 4*: Let $S = (X, U, \to, Y, H)$ be a system and let $Q$ be an equivalence relation on $X$ such that $(x, x') \in Q$ implies $H(x) = H(x')$. The relation $\Gamma = \{(x, x_Q) \in X \times (X/Q) \mid x \in x_Q\}$ is a bisimulation relation between $S$ and $S/Q$ if and only if $Q$ is a bisimulation relation between $S$ and $S$.

## B. Definition of HDS

In brief, a hybrid dynamical system (HDS) is a dynamical system with both discrete-valued and continuous-valued state variables. For a fixed value of the discrete state, the continuous state of HDS evolves as the state of a continuous dynamical system described by ordinary differential equations. There are several definitions of HDS in the literature. Probably the most general one is the "open hybrid automaton" in [6] which is strongly influenced by inputs. It is not a big difficulty to control an HDS with a rich set of inputs and sufficient actuation. To see this, consider the case in which for each discrete state value the corresponding continuous system is controllable. Then the HDS control problem may be divided into continuous system control problems during the intervals with fixed discrete state. By controlling the HDS in this way, any combination of the discrete and continuous states of the HDS is reachable. However, in practical HDS's there are input restrictions with respect to the general case. A very usual restriction is to limit the input to influence only the discrete state directly. This happens for example when the input is itself discrete valued (like a two-state ON/OFF valve in a piping). This work is based on the HDS definition from [13] in the following which considers such an input restriction.

*Definition 5*: A Hybrid Dynamical System $\Sigma$ is a quintuple $(S, \{In_x\}_{x \in X}, \{Gu_t\}_{t \in \rightarrow}, \{Re_t\}_{t \in \rightarrow}, \{f_x\}_{x \in X})$ consisting of
- a finite transition system $S = (X, U, \rightarrow)$;
- a non-empty set $In_x \subseteq \mathbb{R}^n$ for each $x \in X$ denoted as invariant set of $x$;
- a guard set $\emptyset \neq Gu_{(x,u,x')} \subseteq In_x$ for each $(x,u,x') \in \rightarrow$;
- a reset function $Re_{(x,u,x')} : In_x \rightarrow In_{x'}$ for each $(x,u,x') \in \rightarrow$;
- a smooth vector field $f_x : In_x \rightarrow \mathbb{R}^n$ for each $x \in X$.

The state of an HDS is an ordered pair $(x, \xi)$ with $x \in X$ and $\xi \in \mathbb{R}^n$. Evolution of the state over time are denoted by execution of the HDS which is obtained as follows. If $\xi \in In_x$, then $\xi$ can evolve with time $t \in \mathbb{R}$ according to the ordinary differential equation $\frac{d}{dt}\xi(t) = f_x(\xi(t))$. This type of evolution is denoted as a *flow* in $In_x$ (smoothness of the vector fields result in the existence and uniqueness of flows). If $\xi \in Gu_{(x,u,x')}$ at time $t \in \mathbb{R}$ then $(x, \xi)$ may *jump* to $(x', Re_{(x,u,x')}(\xi))$ and then it may continue to flow or jump again. If both flow and jump (or multiple jumps) are possible at a state, then there exists a form of uncertainty. If a flow reaches a point $(x, \xi) \in \partial In_x$ that does not belong to a guard set, then the execution is blocked without being able to proceed in time [6].

### III. Representation of HDS as a Transition System

In this section, we transform the HDS to a transition system by sampling the state variables just before every change in the output values. The concept of equivalence between transition systems is based on having the same external behaviors. A symbolic model has finite valued state and output. Hence, to obtain an equivalent symbolic model for an HDS, we need to define a finite-valued output for the HDS as below.

*Definition 6*: For an HDS $\Sigma = ((X, U, \rightarrow), \{In_x\}_{x \in X}, \{Gu_t\}_{t \in \rightarrow}, \{Re_t\}_{t \in \rightarrow}, \{f_x\}_{x \in X})$, a symbolic output is a map $H_M: \bigcup_{x \in X} \{x\} \times In_x \rightarrow Y_M$ where $Y_M$ is a finite set.

We also need to convert the HDS into a transition system by sampling its output (at some time instants) in order to be able to compare its external behaviors with those of the symbolic model. This is carried out in [13] by introducing the notion of transition system associated with an HDS. In this work, we make the following assumption.

*Assumption 7*: It is assumed that the output of HDS is a function of the discrete state such that for every $x \in X$, $\xi, \xi' \in \mathbb{R}^n$ we have $H_M(x,\xi) = H_M(x,\xi')$.

*Remark 8*: Assumption 7 is not restrictive since due to finiteness of $Y_M$ we can always increase the number of discrete states by dividing the invariant sets and defining the borders between the divided sets as guards such that the output can be determined only from the discrete state. The details are avoided due to the space limits.

Assumption 7 ensures that the output is constant during a flow which enables us to sample the HDS output only at jumps. This results in the smaller transition system in Definition 9 in the following which is more appropriate for computations with respect to the one in [13].

*Definition 9*: For an HDS $\Sigma = ((X, U, \rightarrow), \{In_x\}_{x \in X}, \{Gu_T\}_{T \in \rightarrow}, \{Re_T\}_{T \in \rightarrow}, \{f_x\}_{x \in X})$, with symbolic output map $H_M$ which has a range $Y_M$, the *mapped system* is a transition system $M(\Sigma) = (X_M, U_M, \rightarrow_M, Y_M, H_M)$ with

- $X_M = (X_M^a \cup X_M^b \cup X_M^c) \cap \{(x, \xi) \mid \exists \xi' \in In_x, t > 0 : \xi = \Phi_{f_x}(\xi', t)\}$;

    $X_M^a = \bigcup_{T=(x,u,x') \in \rightarrow} \{x\} \times (Gu_T \cap In_x)$.
    $X_M^b = \{(x, \xi) \mid f_x(\xi) = 0\}$.
    $X_M^c = \{(x, \xi) \mid \xi \in In_x, \theta_{f_x}(\xi) = 0\} \setminus X_M^a$.

- $U_M = U \cup \{*\}$ ;
- $((x, \xi), u, (x', \xi')) \in \to_M$ if either
  1. $x \neq x'$, $T = (x, u, x') \in \to$, $\xi \in \mathrm{Gu}_T$, $\exists\, t > 0 : \Phi_{f_x}(\mathrm{Re}_T(\xi), t) = \xi'$.
  2. $x = x'$, $u = *$, $\xi = \xi' \in \mathrm{In}_x$, $f_x(\xi) = 0$.

The states that cannot be reached from another state are excluded from $X_M$. The state space $X_M$ contains $X_M^a$, the set of points just before jumps on executions of the HDS. The transitions from $X_M^a$ are captured by the first type of transitions in the definition of $\to_M$. Equilibrium states of invariant sets can be regarded as final states that are captured by $X_M^b$. The second type of transitions provides a self transition for each equilibrium to avoid a blocking condition. The points at which an execution is blocked are also captured by $X_M^c$.

Now we can focus on the problem of finding a finite bisimulation for the mapped system which is generally an infinite transition system. To reduce complexities, the class of HDS's considered in this work is restricted as below.

*Assumption 10*: The HDS $\Sigma$ in Definition 5 is assumed to satisfy the following conditions
- for every $x \in X$ the invariant set $\mathrm{In}_x$ is a closed set,
- for every $\xi \in \mathrm{In}_x \cap \mathrm{Gu}_{(x,u,x')}$ we have $\Psi_{f_x}(\xi) = \xi$.

The above assumption eliminates the possibility of jumps when flow is possible which simplifies the analysis in the next sessions.

## IV. BISIMULATION BASED ON BEHAVIORS

In this section, we bisimulate an infinite state transition system $S$ based on its behaviors. The obtained bisimulation has the minimum number of states. In the remaining, it is assume that $S$ which is the mapped system of the HDS is output deterministic.

*Remark 11*: The requirement of output determinism of $S$ is not restrictive in practice. If $S$ is the mapped system of $\Sigma$, It can be easily shown that if the finite transition system of $\Sigma$ (first element in Definition 5) is output deterministic then $S$ is also output deterministic. Even if this is not the case, it is possible to augment the output of $\Sigma$ with additional information such that $S$ becomes output deterministic.

In the remaining, for transition system $S = (R, U, \to, Y, H)$, the set of state sequences of all internal behaviors of length $n$ and all maximal internal behaviors with a length smaller than $n$ that start from $r$ is denoted as $B_n(S, r)$. For every state sequence $b$, the corresponding output sequence is denoted by $H(b)$. Clearly, we have $|b| = |H(b)|$. We define $\mathcal{H}_n(S, r) = H(B_n(S, r))$ and $\mathcal{H}_n(S) = \{\mathcal{H}_n(S, r) \mid r \in R\}$. Also, we define a sequence of partitions of the state space $R$ as below.

$$Q_k(S) = \{(r, r') \in R \times R \mid \mathcal{H}_k(S, r) = \mathcal{H}_k(S, r')\}. \tag{1}$$

Since $S$ is output deterministic, we can define

$$\phi(r, y) = \begin{cases} r' & \exists\, r \xrightarrow{u} r' : H(r') = y \\ \mathcal{O} & \text{otherwise} \end{cases} \tag{2}$$

For $r \in R$, $y \in Y$, the case in which there is no $(r, u, r') \in \to$ such that $H(r') = y$, is indicated as $\phi(r, y) = \mathcal{O}$. If there exists $(r, u, r') \in \to$ with $H(r') = y$, output determinism of $S$ ensures that $r'$ is unique and $\phi$ is well-defined. In the following we present Lemma 12 and Theorem 13 that will provide the means of finding the finite bisimulation.

*Lemma 12*: For a transition system $S = (R, U, \to, Y, H)$, if $|R/Q_k(S)| = |R/Q_{k+1}(S)|$ for some $k > 1$, then we have $Q_m(S) = Q_k(S)$ for every $m > k$.

*Proof*: First, we show that $|R/Q_k(S)| = |R/Q_{k+1}(S)|$ implies $Q_k(S) = Q_{k+1}(S)$. If $r' \in [r]_{Q_{k+1}(S)}$, then according to (1), we have $\mathcal{H}_{k+1}(S, r) = \mathcal{H}_{k+1}(S, r')$. This clearly requires $\mathcal{H}_k(S, r) = \mathcal{H}_k(S, r')$ which means $r' \in [r]_{Q_k(S)}$ and we have

$$[r]_{Q_{k+1}(S)} \subseteq [r]_{Q_k(S)} \quad \forall\, r \in R \tag{3}$$

Since the equivalence classes are mutually disjoint, the above relation implies that every equivalence class of $Q_k(S)$ is split into one or more equivalence classes of $Q_{k+1}(S)$. If $[r]_{Q_k(S)} \neq [r]_{Q_{k+1}(S)}$ for some $r \in R$ then $[r]_{Q_k(S)}$ is split into more than one equivalence classes of $Q_{k+1}(S)$ resulting in $|R/Q_k(S)| < |R/Q_{k+1}(S)|$. Hence, if $|R/Q_k(S)| = |R/Q_{k+1}(S)|$, then for every $r \in R$ we must have $[r]_{Q_{k+1}(S)} = [r]_{Q_k(S)}$ or $Q_k(S) = Q_{k+1}(S)$.

The fact that $Q_m(S) = Q_k(S)$ for $m > k$, is proved by induction if we show that $Q_k(S) = Q_{k+1}(S)$ implies $Q_{k+1}(S) = Q_{k+2}(S)$. To show this, we assume $Q_k(S) = Q_{k+1}(S)$ and $Q_{k+1}(S) \neq Q_{k+2}(S)$, then we show that a contradiction occurs. According to $Q_{k+1}(S) \neq Q_{k+2}(S)$, there exist $r, r' \in R$ such that (i) $\mathcal{H}_{k+1}(S,r) = \mathcal{H}_{k+1}(S,r')$ and (ii) $\mathcal{H}_{k+2}(S,r) \neq \mathcal{H}_{k+2}(S,r')$. There must exist $y \in Y$ such that $\bar{r} = \phi(r,y) \neq \boldsymbol{O}$ and $\bar{r}' = \phi(r',y) \neq \boldsymbol{O}$, otherwise (ii) cannot hold. Hence, we can rewrite (i) and (ii) respectively as (iii) $\mathcal{H}_k(S,\bar{r}) = \mathcal{H}_k(S,\bar{r}')$ and (vi) $\mathcal{H}_{k+1}(S,\bar{r}) \neq \mathcal{H}_{k+1}(S,\bar{r}')$. But, (iii) together with $Q_k(S) = Q_{k+1}(S)$ imply that $\mathcal{H}_{k+1}(S,\bar{r}) = \mathcal{H}_{k+1}(S,\bar{r}')$ which contradicts with (iv). □

*Theorem 13*: For an output deterministic transition system $S = (R, U, \rightarrow, Y, H)$, the transition system which is bisimilar to $S$ and has the minimum number of states is $S/Q_\infty(S)$ with $Q_\infty(S)$ defined according to (1).

*Proof*: According to Theorem 4, first we need to show that $Q_\infty(S)$ is a bisimulation relation from $S$ to $S$. For every $(r,r') \in Q_\infty(S)$ we have $\mathcal{H}_\infty(S,r) = \mathcal{H}_\infty(S,r')$ which gives $H(r) = H(r')$ and fulfills the first assertion of Definition 2. To prove the remaining assertions, consider $\bar{r} \in R$ such that $(r,u,\bar{r}) \in \rightarrow$ is a transition from $r$. If $\bar{b} \in \mathcal{H}_\infty(S,\bar{r})$, then there exists $b \in \mathcal{H}_\infty(S,r)$ such that $b_0 = H(r)$, $b_{1:|b|} = \bar{b}$. Since $\mathcal{H}_\infty(S,r) = \mathcal{H}_\infty(S,r')$, we also have $b \in \mathcal{H}_\infty(S,r')$ which means that there exists $(r',u',\bar{r}') \in \rightarrow$ with $b_{1:|b|} = \bar{b} \in \mathcal{H}_\infty(S,\bar{r}')$ and $H(\bar{r}') = b_1 = H(\bar{r})$. Output determinism of $S$ requires that $\bar{r}'$ is the unique state that gives the output $H(\bar{r}') = H(\bar{r})$. Therefore, for every $\bar{b} \in \mathcal{H}_\infty(S,\bar{r})$ we also have $\bar{b} \in \mathcal{H}_\infty(S,\bar{r}')$ or equivalently $\mathcal{H}_\infty(S,\bar{r}) \subseteq \mathcal{H}_\infty(S,\bar{r}')$. In the same way have the converse and thus equity of the two sets of behaviors. Hence, for every $(r,r') \in Q_\infty(S)$ and $(r,u,\bar{r}) \in \rightarrow$, there exists $(r',u',\bar{r}') \in \rightarrow$ such that $\mathcal{H}_\infty(S,\bar{r}) = \mathcal{H}_\infty(S,\bar{r}')$ or equivalently $(\bar{r},\bar{r}') \in Q_\infty(S)$. This proves that $Q_\infty(S)$ is a bisimulation relation from $S$ to $S$ according to Definition 2.

To show that $S/Q_\infty(S)$ has the minimum number of states, we need to show that the bisimulation relation from $S$ to $S$ which has the minimum number of equivalence classes is $Q_\infty(S)$. According to Proposition 14 in the following, if $Q'$ is a bisimulation relation from $S$ to $S$, then $(r,r') \in Q'$ implies that $(r,r') \in Q_\infty(S)$. It means, any equivalence class of $Q'$ is a subset of an equivalence class of $Q_\infty(S)$. Since the equivalence classes of $Q_\infty(S)$ are disjoint we conclude that every equivalence class of $Q_\infty(S)$ is partitioned by one or more equivalence classes of $Q'$. Hence, the number of equivalence classes of $Q_\infty(S)$ is less than that of $Q'$ and $Q_\infty(S)$ has the minimum number of equivalence classes. □

*Proposition 14*: If $Q'$ is a bisimulation relation from $S$ to $S$, then for every $(r,r') \in Q'$ we have $\mathcal{H}_\infty(S,r) = \mathcal{H}_\infty(S,r')$.

*Proof*: It suffices to show that $\mathcal{H}_\infty(S,r) \neq \mathcal{H}_\infty(S,r')$ with $(r,r') \in Q'$ leads to a contradiction. If $\mathcal{H}_\infty(S,r) \neq \mathcal{H}_\infty(S,r')$, then there exists $b \in \mathcal{H}_\infty(S,r)$ such that $b \notin \mathcal{H}_\infty(S,r')$. Since $(r,r') \in Q'$, we have $H(r) = H(r')$ and for every $b' \in \mathcal{H}_\infty(S,r')$ there exists $k \geq 0$ such that $(b_i, b'_i) \in Q'$ for $0 \leq i \leq k$. We assume that $b'$ gives the maximum value of $k$. There are four cases: 1- $k = \infty$, 2- $|b| = |b'| = k$, 3- $|b| > k$, 4- $|b'| > k$. The first two cases imply $b = b'$ which contradicts with $b \notin \mathcal{H}_\infty(S,r')$. The third case with $(b_k, b'_k) \in Q'$ requires the existence of $(b'_k, u', r'') \in \rightarrow$ such that $(b_{k+1}, r'') \in Q'$ and $[b'\ r''] \in \mathcal{H}_\infty(S,r')$ which contradicts with maximality of $k$. The forth case is also handled in the same way as the third case. □

## V. COMPUTATION

According to Lemma 12 and Theorem 13, if we compute the sets of behaviors $\mathcal{H}_k(S, r)$ for every $r \in R$ successively until $|R/Q_k(S)| = |R/Q_{k+1}(S)|$, then we obtain $Q_\infty(S)$ and we can find $S/Q_\infty(S)$ which is the transition system with minimum number of states that is bisimilar with $S$. Evidently, if $S$ has a finite bisimulation then $Q_\infty(S)$ will have a finite number of equivalence classes. However, if a transition system $S = (R, U, \rightarrow, Y, H)$ is the mapped system of an HDS, then the state space $R$ is infinite (and uncountable) and it is not possible to compute the finite length behaviors for every $r \in R$. To solve this problem, first we observe that:

*Observation 15*: The bisimulation of $S$ in Theorem 13 denoted as $S/Q_k(S)$ can be built from $\mathcal{H}_{k+1}(S)$. There is a one to one correspondence between the state space of bisimulation $R/Q_k(S)$ and $\mathcal{H}_k(S)$. The transitions between elements of $R/Q_k(S)$ can be determined from $\mathcal{H}_k(S)$ and $\mathcal{H}_{k+1}(S)$. There is a transition between two elements of $R/Q_k(S)$ that correspond to $h_1, h_2 \in \mathcal{H}_k(S)$, if there is $h_3 \in \mathcal{H}_{k+1}(S)$ and $y \in Y$ such that $h_1 = \{b \mid \exists b' \in h_3 : b = b'_{0:\max\{k,|b'|\}}\}$, $h_2 = \{b \mid \exists b' \in h_3 : b = b'_{1:|b'|}, b'_1 = y\}$.

If $S$ has a finite bisimulation, then $Q_k(S)$ must have a finite set of equivalence classes (Lemma 12, Theorems 13 and 4). Therefore, if we select a finite set $\mathfrak{R} \subseteq R$ such that for every equivalence class of $Q_k(S)$ there exists $r \in \mathfrak{R}$, then we have $\mathcal{H}_{k+1}(S) = \{HB_{k+1}(S, \mathfrak{R}) \mid r \in \mathfrak{R}\}$ and we will not need to compute $\mathcal{H}_{k+1}(S,r)$ for every $r \in R$. Assumption

16 in the following ensures that $\mathfrak{R}$ is such a subset of $R$. First, should define a distance between $r, r' \in R$ as in (4) in which $r = (x,\xi)$, $r' = (x',\xi')$, $x, x' \in X$ and $\xi, \xi' \in \mathbb{R}^n$ with $X$ being the set of discrete states of the HDS.

$$d(r,r') = \begin{cases} \|\xi - \xi'\| & x = x' \\ \infty & x \neq x' \end{cases} \tag{4}$$

*Assumption 16*: For the HDS $\Sigma$, it is assumed that a finite state equivalent abstraction exists for the mapped system $S = M(\Sigma)$. Also, denoting the state space of $S$ by $R$, it is assumed that there exists $\eta \in \mathbb{R}^+$ such that for every $m \in R/Q_\infty(S)$ there exists $r \in m$ such that the ball with radius $\eta$ centered at $r$ is inside $m$, i.e. $\{r' \in R: d(r',r) < \eta\} \subseteq m$.

With the above assumption, we can select a grid of points in $\mathbb{R}^n$ for each $x \in X$ with sufficiently small distances between the points to build $\mathfrak{R}$. Then $k$, $\mathcal{H}_k(S)$ nd $\mathcal{H}_{k+1}(S)$ are computed using Algorithm 17 in the following.

*Algorithm 17*:
Input: $S = (R, U, \rightarrow, Y, H)$, $\mathfrak{R} \subset R$.
Output: $k$, $\mathcal{H}_k(S)$, $\mathcal{H}_{k+1}(S)$
1: for each $r \in \mathfrak{R}$
2:      $B_0(S,r) \leftarrow \{r\}$ ;
3:      $\mathcal{H}_0(S,r) \leftarrow \{H(r)\}$ ;
4: end
5: $Q_0(S) \leftarrow \{(r,r') \mid H(r) = H(r')\}$ ;
6: $k \leftarrow -1$ ;
7: repeat
8:      $k \leftarrow k+1$ ;
9:      for each $r \in \mathfrak{R}$
10:         $B_{k+1}(S,r) \leftarrow$ proceed$(B_k(S,r), k)$ ;
11:         $\mathcal{H}_{k+1}(S,r) = H(B_{k+1}(S,r))$ ;
12:      end
13:      $Q_{k+1}(S) \leftarrow \{(r,r') \mid \mathcal{H}_{k+1}(S,r) = \mathcal{H}_{k+1}(S,r')\}$ ;
14: until $|\mathfrak{R}/Q_{k+1}(S)| = |\mathfrak{R}/Q_k(S)|$ end

1: procedure proceed$(B_i, k)$
2: $B_o \leftarrow \{b \in B_i: |b| < k \vee \forall y \in Y : \phi(b_k, y) = \mathcal{O}\}$ ;
3: for each $b \in B_i$, $|b| = k$
4:      for each $y \in Y$, $\phi(b_k, y) \neq \mathcal{O}$
5:         $B_o \leftarrow B_o \cup \{[b \;\; \phi(b_k, y)]\}$ ;
6:      end
7: end
8: return $B_o$ ;

*Remark 18*: Algorithm 17 always terminates because we will have $|\mathfrak{R}/Q_{k+1}(S)| \neq |\mathfrak{R}/Q_k(S)|$ for $k \geq |\mathfrak{R}|$. However, such termination means that the number of points in $\mathfrak{R}$ are not sufficient and we must execute the algorithm with smaller values of $\eta$. If we find a value of $\eta$ such that $|\mathfrak{R}/Q_k(S)|$ does not change by decreasing $\eta$, then we can guess that the value of $\eta$ and the corresponding $\mathfrak{R}$ are suitable and the bisimulation computed according to Observation 15 is correct and has the minimum number of states.

## VI. EXAMPLE

In this section we consider a simple example to visualize the obtained results. We consider a temperature control system composed of area 1 with temperature $T_1$ and area 2 with temperature $T_2$ such that area 2 is enclosed in area 1. A heater that can be either in the ON or OFF states is in direct contact with area 1. Dynamical equations for $T_1$ and $T_2$ are given in (5) in which $u(t) = 1$ when the heater is ON and $u(t) = 0$ when the heater is OFF.

$$\dot{T}_1(t) = -T_1(t) + u(t) \tag{5.1}$$

$$\dot{T}_2(t) = -T_2(t) + T_1(t) \tag{5.2}$$

The objective is to reach $T_1 = T_2 = 0$ or $T_1 = T_2 = 1$ from an initial condition inside $T_1, T_2 \in [0,1]$ without exiting a safe region given by

$$R_S = \{(T_1, T_2) \in [0,1]^2 : |T_1 - T_2| \leq 0.25\}. \tag{6}$$

Base on the definition of safe region, we define a symbolic output with range $Y_S = \{\text{safe, unsafe}\}$ as below.

$$H_S(T_1, T_2) = \begin{cases} \text{safe} & (T_1, T_2) \in R_S \\ \text{unsafe} & \text{otherwise} \end{cases} \tag{7}$$

The state of heater is a discrete state of the system that needs to be refined according to Remark 8 as below such that the output can be determined from the discrete state.

$X = \{\text{OFF\_safe, ON\_safe, OFF\_unsafe, ON\_unsafe}\}$

The invariant sets and guards are obtained as below with $\bar{R}_S = \{(T_1, T_2) \in [0,1]^2 : |T_1 - T_2| \geq 0.25\}$.

$\text{In}_{\text{OFF\_safe}} = \text{In}_{\text{ON\_safe}} = R_S$, $\text{In}_{\text{OFF\_unsafe}} = \text{In}_{\text{ON\_unsafe}} = \bar{R}_S$

According to Assumption 10, jumps are allowed on boundaries of invariant sets. Hence, the guard sets are set to the boundaries and since $\partial R_S = \partial \bar{R}_S$, the guard set for every transition $t$ is obtained as $\text{Gu}_t = \partial R_S = \{(T_1, T_2) \in [0,1]^2 : |T_1 - T_2| = 0.25\}$. The reset functions are also the identity function. We select the discrete state as the output used for bisimulation such that $Y_M = X$.

With the above HDS elements, we can compute the mapped system from Definition 9, select a set points in the state space of the mapped system and apply Algorithm 17. The set of 64 points selected on each of the state space of the mapped system are shown in Figure 1. Each of the four segments is depicted on a separate plot and the related discrete states are indicated in rectangles. The points are distributed such that $\eta$ in Assumption 16 is $0.05 \times \sqrt{2}$. The partitions with same behaviors can be recognized by the different symbols used for locating the points in each partition. The partition numbers are also placed beside each group of points in the same partition.

The total number of partitions is 12 which does not increase by decreasing $\eta$. The resulting quotient system which is the obtained bisimulation can be illustrated as the graph of in Figure 2. There are two transitions from each state in Figure 2 associated with ON or OFF control commands to the heater. The control commands are not indicated in the figure for clarity, but they can be determined from the destination of each transition. It is remarkable that the approximate bisimulation method (with $\eta = 0.05\sqrt{2}$ and $\tau = 0.2$ in [13]) results in a non-deterministic transition system with 843 states where the number of states will increase if more accuracy is required.

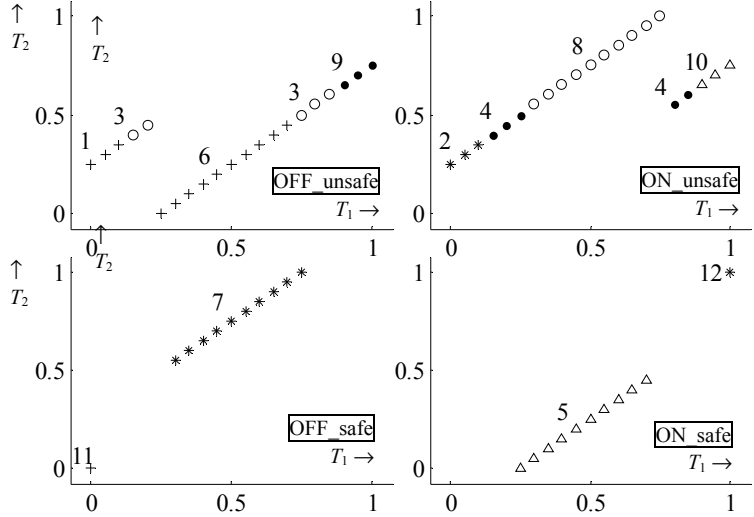

Figure 1. The set of points selected on the state space of the mapped system and their partitioning obtained from Algorithm 17.

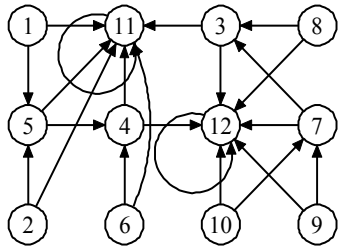

Figure 2. Graph representation of the obtained bisimulation.

## VII. Conclusions

A Methodology was presented for calculating the symbolic model which is bisimilar to a hybrid dynamical system (HDS) and has the minimum number of states. After an analysis of the relationship with behaviors of the system an algorithm was presented to solve the problem. The algorithm may need to be executed for several times to achieve the desired solution.

[1] K.N. Toosi University of Technology, Tehran, Iran.   (e-mail: tavassoli@kntu.ac.ir)